# Real-Time detection, classification and DOA estimation of Unmanned Aerial Vehicle

Konstantinos Polyzos,
Evangelos Dermatas
Dept. of Electrical Engineering &
Computer Technology
University of Patras
ece8481@upnet.gr
dermatas@upatras.gr

## ABSTRACT

The present work deals with a new passive system for real-time detection, classification and direction of arrival estimator of Unmanned Aerial Vehicles (UAVs). The proposed system composed of a very low cost hardware components, comprises two different arrays of three or six-microphones, non-linear amplification and filtering of the analog acoustic signal, avoiding also the saturation effect in case where the UAV is located nearby to the microphones.

Advance array processing methods are used to detect and locate the wide-band sources in the near and far-field including array calibration and energy based beamforming techniques. Moreover, oversampling techniques are adopted to increase the acquired signals accuracy and to also decrease the quantization noise. The classifier is based on the nearest neighbor rule of a normalized Power Spectral Density, the acoustic signature of the UAV spectrum in short periods of time. The low-cost, low-power and high efficiency embedded processor STM32F405RG is used for system implementation. Preliminary experimental results have shown the effectiveness of the proposed approach.

## Introduction

Recently, a great number of organizations or individuals have shown increasing interest in flying Unmanned Aerial Vehicles (UAVs) or drones for surveillance, scientific and reconnaissance purposes or video recordings acquired by UAVs for entertainment [1,2,14,15]. This activity introduces several issues when the UAVs are flying over city centers, official buildings, airport areas, military or nuclear installations. In real-life environment, the detection and tracking them can be a difficult task using traditional techniques. Typical signal processing methods, based on RF data link signatures [3], Radars, cameras [11], proved low reliability methods due to the UAVs RF very low data communication rates or complete navigation autonomy, small vehicles size and low altitude flights. For this reason, there have been copious efforts of trying to find reliable techniques suitable for detection and classification of the acoustic signals generated by the UAVs [7]. An introductory review of the recent advances in detection, tracking and interdiction of UAVs can be found in [12].

Following the well-known and very popular method of linear prediction coding (LPC) [8], which has been used extensively to exploit the time dependence of the signal samples in short periods of time, the UAVs are detected trough LPC classification [4]. In [18], an inexpensive array with dynamically placed microphones is used to detect and locate far-field broadband sources implementing array calibration and beamforming methods. In a short set of experiments, remotely controlled airplanes in an open field are located.

A system composed of a 120-element microphone array and a video camera is presented in [1], and it aims to acoustically detect and track small moving objects, such as drones or ground robots. The acoustic imaging algorithm determines, in real-time, the sound power level coming from all directions, using the phase of the sound signals. Consequently, a beamforming algorithm selectively extracts the sound coming from each tracked sound source. It was determined that drones can be tracked up to 160 to 250 meters, depending on their type and background noise. The experiments were carried out in both, laboratory environments and outdoor conditions.

In [13], an effective feature extraction method for unmanned aerial vehicle (UAV) detection is proposed and verified. The feature vector is composed of a mean and standard deviation of difference value between fundamental frequency and mean variation of their frequency. Qualitative measures based on Fisher scores, show outstanding discrimination of measured UAV signals with low signal to noise ratio (SNR) for these frequency-based features. Detection performance with simulated interference signal is compared to the very popular in speech technology MFCC and shows 37.6% of performance improvement.

One very important subject in acoustics is an accurate estimation of the Direction of Arrival (DOA) of any signal. In this case an array of microphones acquires simultaneous signals, and usually phase information and the far-field hypothesis introduces direction estimators such as Periodogram, MUltiple SIgnal Classification (MUSIC), iterative Sparse Asymptotic Minimum Variance (SAMV) etc. [5].

In this paper a new complete passive method and implementation for real-time detection, classification and DOA estimator of UAVs is presented and assessed . The proposed system composed of a very low-cost commercial hardware components, includes  an array of three to six-microphones, non-linear analog amplification and digital filtering used to avoid saturation effects in the analog signals, in case where the UAV is located near to the microphones, and to increase the detection accuracy when the UAV is flying at long distances. Array processing methods are used to detect and locate the wide band sources in both near and far-field conditions, including array calibration and beamforming techniques, a mandatory part of the system due to the adopted very low-cost hardware implementation. Moreover, oversampling techniques are adopted to increase the acquired signals accuracy, eliminating also the quantization noise introduced by the Analog-to-Digital Conversion process (ADC). The classifier is built on the nearest neighbor rule of a feature set characterizing in detail the UAV spectrum in short periods of time. The features are estimated by performing real-time Fast Fourier Transform (FFT) [6].

In our implementation, the highly optimized DSP software library, provided by the ST Microelectronics, is used in a low-cost CortexM4 microcontroller. Preliminary experiments carried out in three different small size drones, show accurate detection estimation of the quadcopter position and type classification.

### 1. UAV Detection and DOA estimator

Several UAV detection methods have already been proposed and evaluated [13-15,]. In real-life applications three at least microphones offer not only the chance to detect the presence of a drone, but through beamforming and triangulation they can also find its position in 2D-space. Typically, more microphones were used for detection in 3D-space, i.e. estimating the Euclidean distance, azimuth and elevation. In addition, the coherent detection enabled by beamforming also increases the SNR ratio, which consequently increases the detection range, recognition rate and tracking accuracy [19].

In our design, two array configurations were tested, consisting of three and six low-cost condenser omnidirectional miniature microphones. The microphones position can be arbitrary, requiring a calibration procedure to determine their locations each time the system is starting to operate [16]. The calibration process consists of ten short time acoustic pulses generated in arbitrary positions nearby to array. Taking into account the time differences of the pulses arrived at the microphones, their relative positions in the 3D space can be estimated using the multi-dimensional scaling (MDS) method. Detailed presentation of related MDS methods can be found in [19-20].

After array calibration, where both microphones relative position and overall gain parameters are estimated [16,17,18], the system starts the normal operation: In real-time, if the signals power at any microphone exceeds an experimentally derived threshold, the UAV classification process is activated and if an UAV is detected, the corresponding DOA is estimated, in case of three microphones, using the signals energy computed every 200 msecs, i.e. 5 positions per second are estimated. If more accuracy is desired the delay and sum beamformer algorithm between pairs of sensors can be easily applied in the frequency domain and the weighted FFT can be calculated as $H(\omega)=P_{ss}(\omega)/(P_{ss}(\omega)+P_{nn}(\omega))$, where $P_{ss}(\omega)$ is the power spectral density of the signal and $P_{nn}(\omega)$ is the power spectral of the background noise, a convenient formula in cases where the background noise cannot be ignored.

### 2. UAV classification

In the available flash memory of the CortexM4 processor, the data of 32 normalized spectrum data of UAVs signatures can be stored. The classification process uses the nearest neighbor rule to estimate the closest similarity between the recorded normalized spectrum distribution over frequency and time and the library data stored in the micro-controller flash memory. If the minimum weighted-Euclidean distance identifies, for two consequent seconds, the same UAV and it is lower than a pre-defined threshold, the system detects the specific UAV presence.

In the training process, the mean value and the standard deviation at each frequency bin are estimated from real or recorded UAV sounds and are stored in the microcontroller flash memory. Both classification and training procedure can be implemented and run directly in the actual processor in real-time.

### 3. System implementation

*3.1 Analog signal processing*

In our design and implementation, low-cost commercial and popular analog and digital electronic components were used. The hardware consists of three or six condenser microphones, and the amplification/filtering system of each analog channel consists of four operational amplifiers integrated in a single MCP604 chip manufactured by the Microchip. Each amplifier provides a gain bandwidth of 2.8 MHz, low operating current of 230uA, low bias current, high-speed operation, high open-loop gain and rail-to-rail output swing and acceptable, in our application, input voltage noise density of 29 nV/sqrt(Hz). The MCP604 operates with a single supply voltage that can be as low as 2.7V. The amplifiers are connected in series, performing isolation of the microphone signal, high-pass filtering at 80Hz, analog logarithmic amplification, and in the final stage, a linear low-gain amplification stage, extending the analog signal in the range of 0-3.3V, to obtain maximum digitization accuracy for the three ADCs included in the chip of the embedded processor.

Each microphone costs 2.5 Euros and the MCP604, a very popular operational amplifier in low-frequency, low-power and low-cost applications, costs less than 1 Euro.

*3.2 Digital signal processing*

Among the great number of embedded processors, the most recent generation of low-cost, high performance CortexM4 ARM processors (Armv7E-M Harvard architecture) for embedded systems are adopted in our implementation. The Cortex-M4 core features a floating point unit (FPU), single precision which supports all ARM single-precision data processing instructions and data types. The accelerated single precision Floating-Point Unit (FPU) offers up to ten times faster computations over the equivalent integer software library [10]. More specifically, the small-size development board STM32H405 (manufactured by Olimex Ltd [9], at 15 Euro) is used, including the power supply of the analog and digital circuits, and the STM32F405RG ARM Cortex-M4 processor (1.25 DMIPS/MHz). This processor provides a low-cost chip that meets the needs of a microcontroller unit (MCU) with reduced pin count and low-power consumption, while delivering high computational performance and an advanced system response to interrupts, three independent analog-to-digital converters (ADCs) and an efficient Direct Memory Access system. . The development board can be powered from the USB type B connector or using an external power supply at +5V DC.

The processor system clock was set to its maximum frequency of 168MHz and the three ADCs operate at ADCCLK=21MHz and 12 bits resolution (successive approximation converter) which is increased by oversampling implemented by software. Taking into account a conversion rate of 12+3=15 cycles per sample, each ADC has a conversion rate of 21M/15cyc=1.4Msps. Each ADC is sampling the analog signal from one microphone in the three microphones configuration. In the case of six microphones, each ADC is sampling continuously the analog signals of two microphones.

To minimize the CPU workload, the three/six digital signals are transferred to the RAM through a DMA channel. After 3*64 ADC conversions (6*32 ADC conversions for the six microphones configuration) and DMA transfer actions, an interrupt signal is generated by the hardware and the CPU is directed to process the new digital signals, i.e. every 64/1.4Msps=45.71usecs or 21875 times per second. Taking into account the logarithmic amplification in the analog domain, the inverse process performs in the digital domain, i.e. each digital sample is non-linear transferred to the equivalent linear space by computing the corresponding exponent value, and the oversampling process is applied by low-pass filtering and accumulating the 64 consecutive values. This action increases the sampling accuracy by 6 bits, giving an overall sampling accuracy of 12+6=18bits and the sampling frequency drops to 21.875 KHz, which is a reasonable value for both typical UAV power spectrum and microphones bandwidth. The final sampling frequency satisfies the Nyquist criterion for both typical UAV acoustic signals and microphones specifications.

The downsampled signals are stored into RAM and every 2048 samples, (and thus 21.875/2048=10.681ms), the power and the phase distribution over frequency of the three/six signals are estimated continuously in real time using the fast Fourier transform (FFT). The adopted implementation is very accurate, providing frequency accuracy of approximately 10Hz per bit. In the CortexM4 processor, three signals FFT of 2048-samples and the power spectrum estimation are completed within 1.21msecs, which is less than 10% of the processor computing capabilities Typical current consumption for the complete system including analog hardware is 130mA at 3.6V, which is quite satisfactory for battery powered devices.

The software is designed, downloaded (and debugging is also made) when running in the embedded processor using the System Workbench toolchain called SW4STM32, available from www.openstm32.org, which is a free, multi-OS software development environment based on Eclipse, supporting the full range of STM32 microcontrollers and a great number of associated boards. In a laptop running Ubuntu 16.04-LTS OS, the gcc compiler embedded toolchain 5.4, the standard peripheral library V1.8.0 and the DSP highly optimized library Revision: V.1.4.5, provided by the ST microelectronics was installed and used for the implementation of the complete system including peripheral programming, i.e. GPIOs, ADC, DMA, System clocks, USART communication to main computer, interrupt and signal processing functions and power spectrum estimation.

## 4. Power spectral density of 3 different types of UAVs

In Figures 1,2,3, the power spectral density up to 15KHz and a photo of three different types of UAVs, are shown; the Quadcopter DJI P3, Quadcopter CX10, and the Quadcopter Sennheiser MKH 8040. The acoustic signal is sampled at 44.1 KHz.

The spectral density is significantly different between Quadcopters, following two principles: smaller propeller and quadcopter size leads to greater power at higher frequencies and the quadcopters introduce wide-band spectral density.

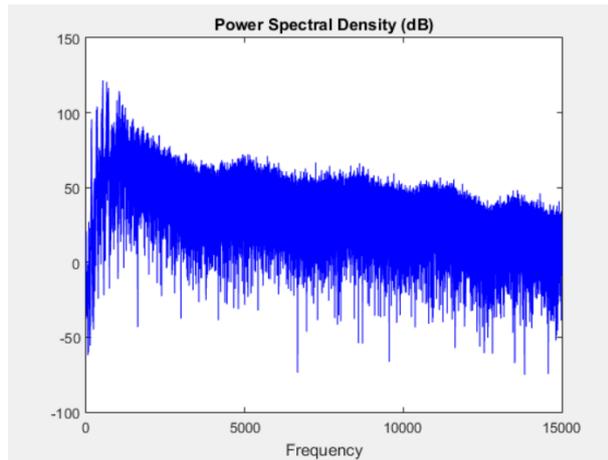
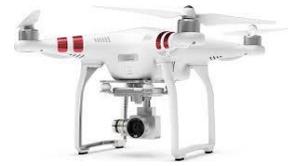

*Figure 1: Power Spectral Density of Quadcopter DJI P3*

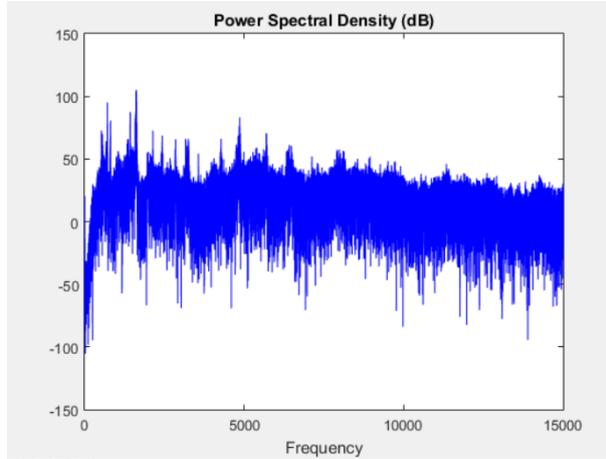
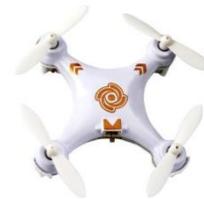

*Figure 2: Power Spectral Density of Quadcopter CX 10*

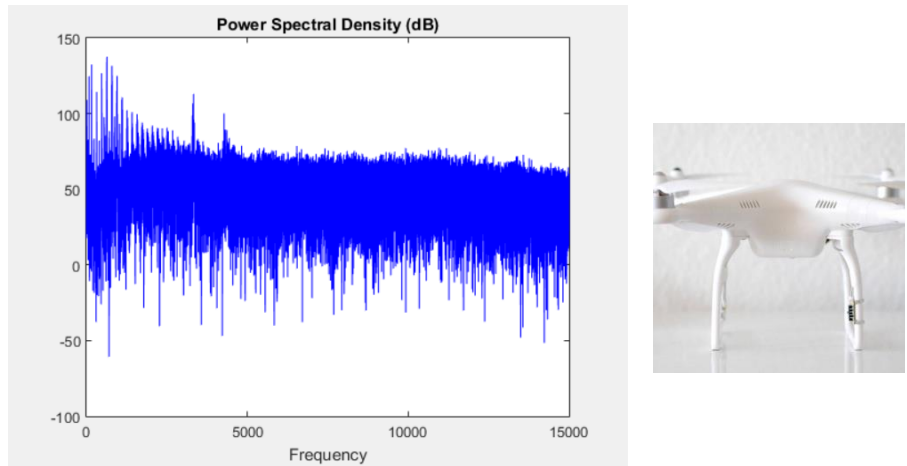

*Figure 3: Power Spectral Density of Sennheiser MKH 8040*

## 5. Conclusion

A complete autonomous low-cost, battery powered acoustic and recognition type system of detection and tracking regarding UAVs, based on a small number of condenser omnidirectional microphones, has been developed and assessed. Detection and tracking algorithms were implemented using an energy-based beamforming algorithm, a weighted distance function for quantitative measure of similarities between the normalized power spectral density of an audio fingerprint database of known quadcopters and the flying UAV. A detailed presentation of the implementation, including both analog and digital processing of microphone signals is given. In preliminary experiments, the detection and the classification module never fail in noise-free environment. The total cost of the complete hardware in end-user prices is less than 30 Euros. Further experiments in more realistic environment will be carried out, improving also the location and tracking accuracy using the MDS framework.

## 6. References

[1] J. Busset, F. Perrodin, P. Wellig, B. Ott, K. Heutschi, T. Rühl, T. Nussbaumer, "Detection and tracking of drones using advanced acoustic cameras" Unmanned/Unattended Sensors and Sensor Networks XI, and Advanced Free-Space Optical Communication Techniques and Applications Proceedings V. 9647, (2015) SPIE Security+Defence, Toulouse, France, 2015

[2] S. Boddhu, M. McCartney, O. Ceccopieri, R. Williams, "A collaborative smartphone sensing platform for detecting and tracking hostile drones", Proceedings Volume 8742, Ground/Air Multisensor Interoperability, Integration, and Networking for Persistent ISR IV, SPIE Defense, Security, and Sensing, Baltimore, Maryland, United States, 2013

[3] M. Peacock, "Detection and control of small civilian UAVs", Bachelor thesis, Edith Cowan University, 2014

[4] J. Vilimek, and L. Burita, " Ways for Copter Drone Acustic Detection" , International Conference on Military Tecnologies (ICMT), 2017

[5] R. Adve, *"Direction of Arrival Estimation"* , Notes, Department of Electrical and Computer Engineering, University of Toronto, 2017

[6] J. Sunu, A. Percus, "Dimensionality reduction for acoustic vehicle classification with spectral embedding", *Sensing and Control (ICNSC)*, IEEE, 2018

[7] L. Hauzenberger, and E. Holmberg Ohlsson, "Drone Detection using Audio Analysis" , Master's Thesis, LTH, Lund University, June 2015

[8] L. Grama, E. Buhus,and C. Rusu, "Acoustic Classification using Linear Predictive Coding for Wildlife Detection Systems" , IEEE, 2017

[9] https://www.olimex.com/Products/ARM/ST/STM32-H405/

[10] https://developer.arm.com/products/processors/cortex-m/cortex-m4

[11] Shuowen Hu, Geoffrey H. Goldman, and Christoph C. Borel-Donohue, "Detection of unmanned aerial vehicles using a visible camera system," Appl. Opt. **56**, B214-B221, 2017

[12] I. Guvenc, F. Koohifar, S. Singh, M. L. Sichitiu and D. Matolak, "Detection, Tracking, and Interdiction for Amateur Drones," in *IEEE Communications Magazine*, vol. 56, no. 4, pp. 75-81, APRIL 2018.

[13] K. Juho, L. Kibae, B. Jinho, and L. Chong Hyun, "Feature Extraction Algorithm for Distant Unmmaned Aerial Vehicle Detection", *Journal of the Institute of Electronics and Information Engineers* Volume 53, Issue 3,  pp.114-123, 2016,

[14] G. C. Birch, J. C. Griffin, and M. K. Erdman, "UAS Detection, Classification, and Neutralization: Market Survey," prepared by Sandia Nat'l. Labs, 2015.

[15] M. Benyamin and G. H. Goldman, "Acoustic Detection and Tracking of a Class I UAS with a Small Tetrahedral Microphone Array," Army Research Lab. tech. rep. (ARL-TR-7086), DTIC Doc., Sept. 2014.

[16] K. Yao, R.E. Hudson, C.W. Reed, D. Chen, & F. Lorenzelli, "Blind Beamforming on a Randomly Distributed Sensor Array System", IEEE Journal on *Selected Areas in Communication*, 16th ed. 1555-1567, 1998

[17] Birchfield S., Subramanya, A. "Microphone Array Position Calibration by Basis-Point Classical Multidimensional Scaling. Speech and Audio Processing", IEEE Transactions on. 13. 1025 – 1034, 2005

[18] S. T. Birchfield, "Geometric microphone array calibration by multidimensional scaling," IEEE International Conference on *Acoustics, Speech, and Signal Processing, 2003. Proceedings. (ICASSP '03).*, Hong Kong, pp. V-157, 2003

[19] E. Case, A. Zelnio and B. Rigling, "Low-Cost Acoustic Array for Small UAV Detection and Tracking," *IEEE National Aerospace and Electronics Conference*, Dayton, OH, pp. 110-113,2008